# A Design Study of a Compact Small-Angle Neutron Scattering Instrument

Markus Bleuel[1,2], Miriam Siebenbürger[3], Peter Böni[4], Gerald J. Schneider[5,6]

[1] NIST Center for Neutron Research, National Institute of Standards and Technology, Gaithersburg, Maryland 20899-8562, United States

[2] Department of Materials Science and Engineering, University of Maryland, College Park, Maryland 20742-2115, USA

[3] Center for Advanced Microstructures and Devices, Louisiana State University, 6980 Jefferson Hwy., Baton Rouge, LA 70806

[4] Physik Department E21, Technische Universität München, D-85748 Garching, Germany

[5] Department of Chemistry, Louisiana State University, Baton Rouge, LA 70820

[6] Department of Physics and Astronomy, Louisiana State University, Baton Rouge, LA 70820

## Abstract

Nanoscale structure determination belongs to one of the crucial tasks in materials science. Small-angle neutron scattering (SANS) is a highly valuable tool to investigate nanostructures. Here, we explore the possibility of a compact SANS instrument to be installed at an individual accelerator based pulsed low-flux source, and discuss applications in structural characterization, education, and training. Monte Carlo simulations of a realistic setup demonstrate the feasibility of such an instrument, with an ideal measurement taking about 7 hours. The anticipated length-scale measurement range is 0.5 nm to 50 nm. The minimalistic design results in an easy to operate, low maintenance instrument, independent of the schedules of large-scale facilities. Size, cost, and maintenance are comparable to laboratory small-angle X-ray scattering (SAXS) instruments, which makes SANS affordable for individual research teams at universities and provides unique chances for education and training. The proposed setup creates the opportunity to measure ultraslow structural changes, at a fraction of the cost of existing SANS instruments, hence enables weekly or monthly repetition rates. The concept can be expanded to an entire suite of instruments, including wide-angle neutron scattering (WANS) and diffraction, neutron reflectometry (NR), neutron radiography, grating interferometry (GI), and prompt gamma activation analysis (PGAA).

# 1 Introduction

Small-angle neutron scattering (SANS) is often the method of choice if contrast needs to be varied to obtain structural information at length-scales ranging from ~ 0.1 nm to ~ 1 µm. Typically, the large scattering length difference between Hydrogen and Deuterium for neutrons is utilized by the soft matter community, and sensitivity of the neutron spin to magnetic structures permits determination of magnetic form factors on that length scales. SANS instruments are used as workhorses to explore the Nanometer regime, there are numerous examples for fundamental and applied science investigated by SANS experiments, like defects in metals, conformations or structure peaks of polymers and biological macromolecules or magnetic phenomena like skyrmions and superconductivity.[1-10] As consequence of the enormous range of potential applications, SANS instrumentation can be found at every major neutron scattering facility, at reactor and spallation sources, are highly oversubscribed and the need for SANS beamtime is intensified by the limited number of large-scale facilities. Additionally, multiple shutdowns of established large-scale neutron centers in the past and near future further limit the availability of SANS.

To provide users with enough neutrons several small-scale accelerator-driven neutron sources (CANS) facilities are planned, with a substantial interest of the community which led to the foundation of the union of accelerator-driven neutron sources (UCANS) in 2010.[11] UCANS include already built facilities like LENS in the US, HUNS, KUANS and RANS in Japan,[12] but also planned projects in Europe like the ESS Bilbao, Sonate (CNRS), and NOVA ERA (JCNS) or the C-CANS in Canada.[13, 14] Typically, these facilities are designed to host around 3-4 instruments. Depending on the size, the investment ranges between 10 and 400 million EUR.[13]

High intensity SANS instruments enable kinetic studies to explore time dependent effects like micellization or changes caused by external fields observed over minutes to several days, only limited by the number of days assigned to a specific user.[15, 16] While it appears to be simple to propose measurements to identify structural changes over weeks and months, and then to measure over multiple reactor cycles, there seem to be little to none of such experiments reported in literature.[17 18] At least they represent a very minor fraction on regular schedules. Logistics of shipping or storing samples, together with regular maintenance and upgrades of instruments add to the constraints. In other words, the working schedule for spallation and reactor neutron sources may be subject to changes in response to evolving projects, operational, and outside needs. Interesting time dependent effects may occur during a shutdown, which renders long-term studies practically non-feasible at existing SANS as the timescale is determined by the samples and is often incompatible with the operation of a large-scale scattering facility. There are many examples of slow changes in a variety of systems, including but not limited to concrete, plastics, batteries, and fuel cells.[19-23] The underlying processes may be catalyzed by UV, chemicals, temperature, and mechanical forces, and respective structural changes may include slow decomposition, (bio)corrosion, erosion phenomena, directed nano-structure assembly, slow growth and deposition processes, or the slow time-dependent loss of magnetic information, cavity and crack formation or agglomeration, to name only a few. So far there is little possibility to gain information on such processes with SANS, which limits our progress in the fundamental understanding of materials.

One approach to tackle this challenge is to build instruments that are independent of the scheduling constraints of reactor or spallation sources, for example the here proposed compact SANS. The existence of such an instrument would benefit measurements of slow processes, but also allow for an easier access of SANS experiments for a broader scientific community, increasing scientific progress and adding opportunities for additional training and education, since in the current situation the tight user programs at existing SANS beamlines put strict limits on beamtime available for educational purpose. Here, we introduce a lower flux source than typical compact sources would offer but paired with only a single experiment, which re-gains flexibility.

Here, we show that recent advancements in neutron generator technology and their continuously increasing flux enables a technical realization of a source capable to produce continuous and pulsed neutrons. Such a versatility in repetition frequency and neutron pulse width allows optimization of the neutron flux to the experimental needs down to specific requirements of a particular sample and can also be used to mirror beam characteristics of existing neutron sources at large-scale facilities to develop and test neutron instrument components or sample environment prototypes under realistic conditions.

The compact SANS design aims to be primarily operated at institutions like universities and laboratories that may not have the infrastructure, space, and staff to operate large scale instrumentation. It is optimized under the joint constraints of safety, durability, easy maintenance, simplicity, and cost of operation. It creates numerous opportunities for general education courses, training of the next generation workforce in neutron science, especially researchers at the early stage of their scientific careers, including undergraduate and graduate students, postdocs, and faculty but also scientists and engineers from companies and research centers. SANS separated from the tight schedule of a large-scale facility allows user training at universities, saving highly precious neutron instrument time.

Easier accessibility is also one important pre-requirement for a larger community participating in instrument and sample environment development as the technical optimization associated with commissioning of any new soft and hardware for instrument components like gratings, neutron lenses or specialized sample chambers highly benefits from local testing.

For scientists concentrating on scattering experiments, conducting preliminary studies, including contrast variation, testing of contrast matching conditions, initial structure examination of new samples, will accelerate the pace of writing competitive proposals, foster success of follow-up experiments at large-scale facilities, as exemplified by the SAXS community. As in their case, a compact SANS can be powerful enough to lower the workload of existing SANS instruments and introduces opportunities to conduct research on slow structural changes that require regular access over weeks, month or even years.

The current paper concentrates on presenting a design for a compact SANS instrument, but the idea can be applied to other concepts like wide-angle neutron scattering (WANS), neutron reflectometry (NR), neutron radiography, grating interferometry (GI), prompt gamma activation analysis (PGAA), and positron annihilation spectroscopy (PALS), which can even be implemented within the same instrument platform. In general, neutron spectroscopy fits also well in this anticipated suite of neutron scattering experiments.

In the following sections the outline of a very compact generator based ToF-SANS instrument is shown, focusing on an affordable and very simple technical realization.

## 2 Theory

This section summarizes relevant equations useful for the design of a time-of-flight (ToF) small-angle neutron scattering (SANS) instrument. SANS experiments measure the scattered intensity as a function of the scattering angle, $2\theta$, or the related momentum transfer

$$Q = \frac{4\pi}{\lambda} \sin(\theta) \qquad \text{Eq. 1}$$

with the wavelength, $\lambda$, of the incident neutrons. $\lambda$ can be calculated:[24]

$$\lambda = \frac{h}{v \cdot m_N} \qquad \text{Eq. 2}$$

with the Planck constant $h \approx 6.626 \cdot 10^{-34}$ Js, the velocity $v$ and the neutron mass $m_N \approx 1.675 \cdot 10^{-27}$ kg. The wavelength spectrum of thermal neutrons is described by the Maxwell-Boltzmann distribution[25]

$$f(\lambda) = 2 \cdot \frac{\lambda_T^4}{\lambda^5} \cdot \exp\left(\frac{-\lambda_T^2}{\lambda^2}\right) \qquad \text{Eq. 3}$$

with the definition $\lambda_T^2 = \frac{h^2}{2 m_N \cdot k_B \cdot T}$, with the Boltzmann factor $k_B \approx 1.38 \cdot 10^{-38}$ J/K and the moderator temperature, $T$.[26] The factor 2 results from the normalization of the Maxwell-Boltzmann distribution $\int_0^\infty f(\lambda) d\lambda = 1$.

The $Q$-resolution depends on the wavelength resolution $\sigma_\lambda$ and the angular/geometric resolution $\sigma_\theta$. The compact SANS instrument is designed for a pulsed source. The wavelength resolution for the corresponding wavelength $\sigma_\lambda(\lambda) = \Delta\lambda/\lambda$ is given by the ratio of the neutron pulse or opening time of the chopper $\Delta t$ and the time of the neutron to fly the distance to the detector ToF ($\Delta t$/ToF). Additionally, the time difference due to the finite thickness of the moderator, $\Delta L$, compared to the instrument length $L$ has to be considered as well.[27]

$$\frac{\Delta Q}{Q} = \sqrt{\sigma_\lambda^2 + \sigma_\theta^2} = \sqrt{\left(\frac{\Delta t}{ToF}\right)^2 + \left(\frac{\Delta L}{L}\right)^2 + (\cot\theta \Delta\theta)^2} \qquad \text{Eq. 4}$$

The value of $\sigma_\theta$ depends on the collimation, the pixel size, the beam size, the effects of gravity and the length of the instrument. In the case of small angles, where $\sin(\theta) \approx \theta$, the last term of Eq. 4, $(\cot\theta\Delta\theta)^2$, can be simplified to $(\Delta\theta/\theta)^2$. Neglecting gravity which is reasonable for short instruments, the geometric resolution $\sigma_\theta$ for a pinhole setup according to Mildner & Carpenter,[28, 29] is given by

$$\sigma_\theta^2 = \frac{1}{12}\left[3\left(\frac{R_C}{L_C}\right)^2 + 3R_S^2\left(\frac{1}{L_C} + \frac{1}{L_{SD}}\right)^2 + \left(\frac{\Delta R}{L_{SD}}\right)^2\right] \qquad \text{Eq. 5}$$

with the circular apertures of radii of collimation $R_C$ and sample $R_S$, the length of the collimation $L_C$, the sample to detector distance $L_{SD}$, and the detector radial bin width $\Delta R$ at the radius $R$ on the detector for the corresponding $|Q|$.

For a compact instrument using a short Soller collimation focusing the neutrons on one spot at the detector the distance from the exit of the collimation to the sample position denoted as $L_{CS}$ has to be considered. In this case,[30, 31]

$$\sigma_\theta^2 = \frac{1}{12}\left[\frac{\left(w_{1_x}^2 + w_{1_y}^2\right)(L_{CS} + L_{SD})^2}{2L_C^2 L_{SD}^2} + \frac{\left(w_{2_x}^2 + w_{2_y}^2\right)(L_C + L_{CS} + L_{SD})^2}{2L_C^2 L_{SD}^2} + \frac{w_{\text{pix}_x}^2 + w_{\text{pix}_y}^2}{L_{SD}^2}\right] \qquad \text{Eq. 6}$$

with the width of the collimator opening $w_{1_x}$ in horizontal and $w_{1_y}$ in vertical direction and the sample aperture $w_{2_x}$ and $w_{2_y}$ and the pixel size $w_{\text{pix}_x}$ and $w_{\text{pix}_y}$, respectively.

Slit smearing is a common approach to increase the intensity and often used in the X-ray scattering community and several procedures for data analysis have been developed.[32, 33] Collimating the neutron beam in one direction only increases the intensity, but leads to slit smearing in one dimension with high azimuthal resolution. For parameters of the instrument discussed in detail in the next chapters, the scattered intensity is increased by a factor of $\approx 60$ compared to a similar collimation in both dimensions, as the direct beam spot is $\approx 30$ times larger in the unfocussed direction and a two times bigger sample volume can be used. **Figure 1** illustrates that one-dimensional slit smearing distorts the scattered intensity profile and reduces the visibility of valuable details in the scattering curve.

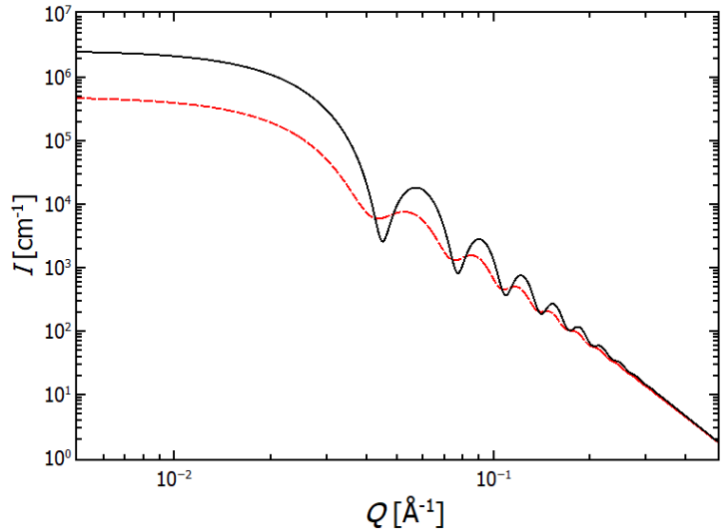

**Figure 1:** Modeled scattering curve of 10 nm spheres with 5% polydispersity. Additionally, the black line represents $\Delta Q/Q = 10\%$ pinhole resolution. The red dashed line is calculated with a constant $0.1 \ \text{Å}^{-1}$ slit resolution instead.

There are two ways to treat slit-smeared data: **(1)** Either trying to remove the effect from the data by the desmearing method of Lake,[32, 33] which has the potential drawback that the results might overinflate backgrounds, create artefacts and the result of this fitting process is not necessarily unique.[34] The advantage of this method is that the output can be directly compared to other SANS and SAXS data and Fourier transformed into real space information. **(2)** The other method convolutes the model with the instrumental resolution during the data fitting process.

$$I_s(Q) \approx \int_0^{\Delta Q_u} du \, I\sqrt{(Q_x^2 + u^2)} \qquad \text{Eq. 7}$$

with $\Delta Q_u$ the slit-height in momentum space along the elevation angle.

## 3 Challenges and Design Choices of a Compact SANS

This section outlines the principles of a compact minimalistic SANS instrument. **Figure 2** illustrates basic components of a scattering instrument: neutron source, moderator, collimator, sample position, flight path, and detector.

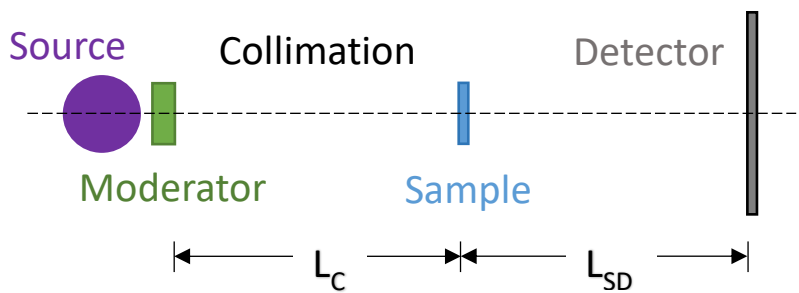

**Figure 2:** Schematic of a typical SANS instrument (accelerator driven fast neutron source in purple, moderator in green, sample in blue, and detector in grey). The length of the collimation is $L_C$ and the sample to detector distance is

For the sake of cost, simplicity and durability moveable components are avoided, also reducing maintenance which implies that the neutron detector is installed at a permanent position, and no choppers are used. A pulsed source is used to scan a range of wavelength and options to remove long wavelength neutrons to avoid frame overlap need to be considered.

The compact SANS instrument is shorter (length ≈ 2 m) than typical existing instruments (length ≈ 10-40 m).[16] The shorter flight path implies less requirements for vacuum to prevent significant scattering along the flight path (0.1 – 1 hPa), and gravitational effects on neutrons are less important, as even 10 Å neutrons decent less than 1 mm over that distance.

Here, we prefer vacuum, because filling the flight chamber with low absorption gases like Argon would require considering neutron activation, which would make it more difficult for potential university laboratory operators. As cosmic radiation incidents, local photon sources, like light bulbs, and potentially other sources may increase the background, shielding will be required. Shielding of the instrument, combined with an interlock system will prevent access of the area when the beamline is active for operational and environmental safety.

Typical length-scales measured by SANS require cold neutrons with wavelengths up to several tens of Å. As illustrated in in **Figure 3**, assuming a Maxwell-Boltzmann distribution (Eq. 3) for a moderator at room temperature ($T = 298$ K) leads to a maximum at $\lambda \approx 1.1$ Å with an amplitude of $\approx 90\%$, and only $\approx 0.02\%$ left at 10 Å, showing the importance of the moderator temperature. For comparison, at a wavelength of 0.5 Å the amplitude is $\approx 0.19\%$ of the total neutron flux. The use of a colder moderator can increase the cold neutron flux by about an order of magnitude.

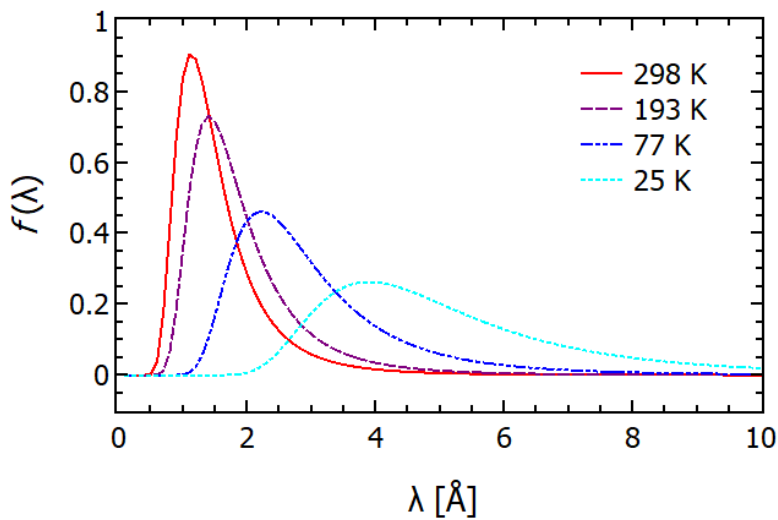

**Figure 3**: Wavelength dependent Maxwell-Boltzmann distribution for different temperatures, calculated using Eq. 3.

While the practical wavelength minimum is determined by the moderation temperature, the maximum wavelength is determined by the frame overlap and can be selected by the time-of-flight, and the instrument length, $L = L_\text{C} + L_\text{SD}$. **Figure 5** shows the achievable maximal wavelength and the related minimum momentum transfer, $Q$, for different $L$ at a neutron source frequency of $f = 200$ Hz. For this example, a convenient length would be 2 m, and with a pulse repetition time of 5 ms or $\lambda_\text{max} \approx 10$ Å neutrons can be used without frame overlap (see **Figure 4**). For comparison, at $f = 50$ Hz, the time-of-flight can reach up to 20 ms allowing for $\lambda_{max} \approx 40$ Å. For practical purposes a total length of 2 m is a rational choice, which fits very well to a typical lab space. For a room temperature moderator with low intensity for colder neutrons, frame overlap can most likely be neglected for most experiments. An option to install a frame overlap mirror (FOM) to remove longer wavelength as a source of background is discussed in chapter 4.2.

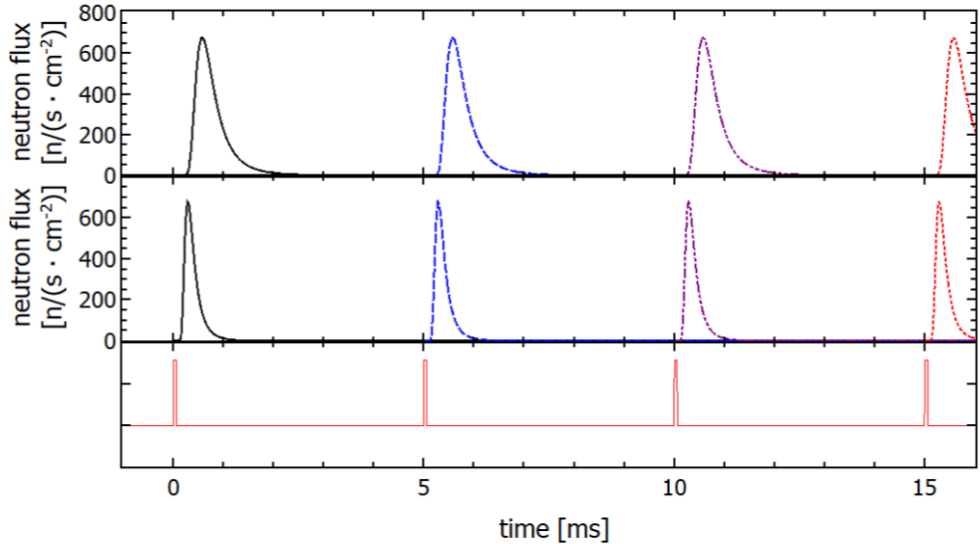

**Figure 4:** (Bottom panel): time structure of the neutron pulses ($t_{pulse} = 50$ µs) in the generator at a frequency of 200 Hz and the moderator at room temperature; (middle panel): calculated time dependent neutron flux at 1 m; (top panel): calculated time dependent neutron flux at 2 m.

At a given wavelength, the lowest $Q$ is limited by the minimum angle, which depends on the instrument length, pixel size, and blurring of the direct beam. **Figure 6** illustrates the increase of the wavelength uncertainty $\sigma_\lambda = \Delta\lambda/\lambda$ with decreasing instrument length $L$. In addition, the geometric uncertainty $\sigma_\theta$ decreases with length $L$ (see **Figure 6**). Hence, SANS instruments have a practical minimum length of $L = L_C + L_{SD} = 2$ m, which means $L_{SD} = L_C = 1$ m. For short instruments, the thickness of the moderator ($\Delta L$) adds a significant wavelength uncertainty (see Eq. 4). For a 2 m instrument with a moderator of 50 mm thickness, the uncertainty would be already $\Delta L/L \approx 2.5\%$.

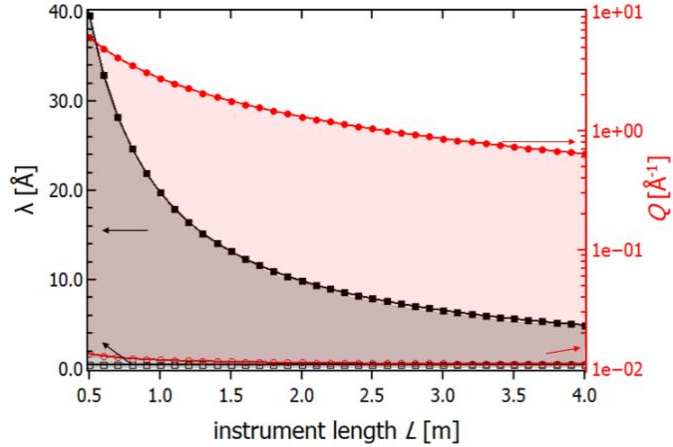

**Figure 5**: Dependence of wavelength and momentum transfer on instrument length. Left axis: $\lambda_{min}$ (black empty squares), defined by the moderator temperature, and $\lambda_{max}$ (black filled squares), determined by the instrument length. Right axis: Corresponding $Q$-range, with $Q_{min}$ (red empty circles) and $Q_{max}$ (red filled circles). The calculation assumes $f = 200$ Hz, a 10 cm detector radius around the beam center , with $Q_{min}$ depending on the size of the beamstop and the pixel resolution (see chapter 4.2).

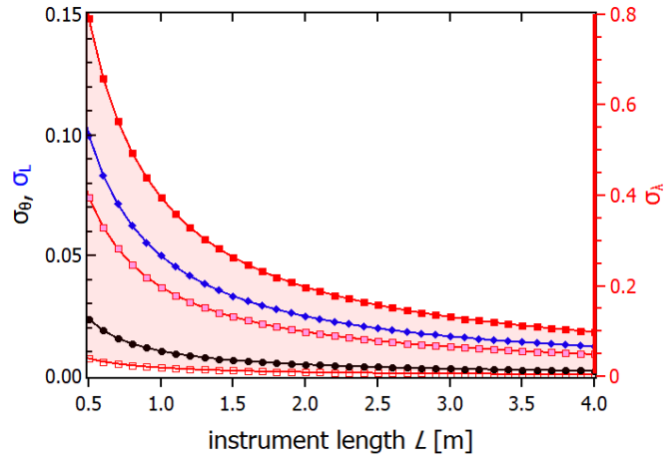

**Figure 6:** The geometric uncertainty $\sigma_\theta$ calculated by Eq. 6 *vs.* the instrument length $L$ for a pulse length of 50 μs is shown in black circles, with $w_{1_x} = 10$ mm, $w_{2_x} = 5$ mm, $w_{1_y} = w_{2_y} = 100$ mm, $w_{pix_x} = w_{pix_y} = 6$ mm $L_C = 1$ m and $L_{SD} = 0.95$ m. The uncertainty in length $\sigma_L$ is depicted in blue. The wavelength uncertainty $\sigma_\lambda = \Delta\lambda/\lambda$ is depicted in red squares (the filled symbols represent calculations for $\lambda = 0.5$ Å and hollow symbols for $\lambda = 10$ Å. Additionally, the $\sigma_\lambda$ for $\lambda = 1$ Å is shown in light pink squares).

The main results of this design section are: (a) For a given 1 m fixed sample to detector distance, a wavelength range from 0.5 to 10 Å and a $Q$-range from 0.012 to 2.6 Å$^{-1}$, with a resolution of $\Delta Q/Q \sim 10\% - 20\%$, can be accomplished. The corresponding length-scale regime extending from 1 Å to 500 Å is highly useful to study the examples provided in the introduction, such as suspensions, vesicles and gels, magnetic nanoparticles, pores, or grains in solid samples, without imposing hardware or cost constraints that would be difficult to accomplish at a university.

## 4 Technical Realization

In this section, the technical realization of the concept presented in section 3 is described, using commercially available items, leading to an affordable and useful instrument for university laboratories, requiring only little training, and offers maximum safety. As **Figure 7** illustrates, the basic setup includes an accelerator driven neutron source with moderator, collimator, sample, and detectors. The minimalistic technical realization includes vacuum flight path to reduce absorption and shielding to lower the background and increasing environmental safety.

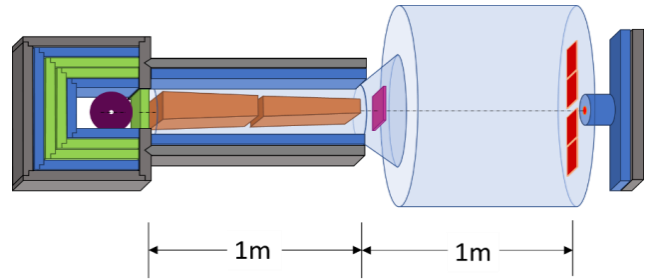

**Figure 7:** Isomeric view of the design: the shielding is color coded with green for polyethylene (PE), blue for boron enriched PE (BPE) and gray for lead (Pb). The source is depicted in purple, the collimators in orange, the sample in magenta, the vacuum tube in transparent light blue and the detectors (for scattering and for transmission) in red. On the right, the beam dump is indicated in blue and gray for BPE and Pb, respectively.

*4.1 Neutron Source and Moderator*

Small accelerator driven fusion based neutron generators are potential sources of neutrons at universities.[35] Currently, various commercial neutron generators are available using either deuteron-deuteron (DD) or deuteron-tritium (DT) reactions to produce neutrons with an energy of 2.5 or 14.1 MeV, respectively.[35] Due to intense research these generators have matured and have been successfully used for imaging, activation experiments, materials analysis, and medical isotope production.[35, 36]

When it comes to SANS, DD generators are advantageous, because neutrons of 2.5 MeV require less moderation, and the lower background intensity at the source reduces shielding required. In addition, the avoidance of tritium reduces the need of a tight sealing of the generator.

In summary, a compact SANS instrument could take advantage of moderated neutrons from a DD fusion reaction[37]

$$^{2}_{1}H + ^{2}_{1}H \rightarrow ^{2}_{2}He + ^{1}_{0}n + 2.45 \text{ MeV} \qquad \text{Eq. 8}$$

Reliable commercial sources using this reaction are available from various suppliers. For example, the DD109M manufactured by Adelphi Technology produces thermal neutrons with a flux of the order of $10^{8} \text{n cm}^{-1}\text{s}^{-1}$.[38] This generator is based on a central target cavity, surrounded by four radial deuteron ion sources. Electron cyclotron resonance (ECR) accelerates deuteron ions to reach 120 keV on a titanium target. Fast neutrons of energy 2.45 MeV are produced with a subsequent thermalization by a polyethylene (PE) moderator. Hereafter, we concentrate on neutron scattering omitting details on commercially available neutron sources that can be found elsewhere.[35, 38]

The fusion reaction (Eq. 8) can be used to produce a continuous or a pulsed beam of neutrons. The Adelphi generator DD109 can be operated with a minimum pulse width of $\geq 0.05$ ms, and a repetition frequency, $f_n \leq 1000$ kHz.[39] Within these specifications, the intensity in each pulse is independent from the repetition rate. The generator pulse repetition frequency and pulse width are flexible. Thus, the operator can easily switch between higher frequencies with a smaller $Q$-range to gain intensity for faster contrast matching tests, as well as lower source frequencies to access a lower $Q_{min}$ in long-term experiments, where the stability of the setup is more important than the duration of an individual scan. The ability to run in a continuous mode allows for white beam experiments at maximum intensity but limited $Q$-resolution.

The compact neutron generator with subsequent polyethylene (PE) moderator at ambient temperature produces a beam with a useable cross sectional area greater than 100 mm × 100 mm, see **Figure 8**.[40]

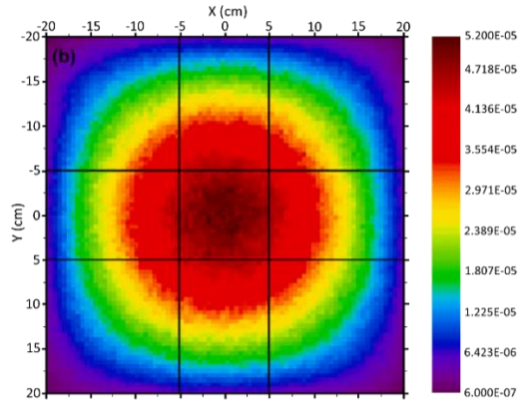

**Figure 8:** 2D intensity distribution on the detector from a neutron generator.[40] The solid lines encompass the area of 100 mm × 100 mm used for our considerations.

With the intensity per pulse independent on the repetition rate, the number of pulses per second determines the intensity, hence the repetition rate should be maximized, but the finite ToF of neutrons causes frame overlap. However, due to the much higher flux of shorter wavelength neutrons, we expect a minor background contribution for SANS experiments, which can also be eliminated by a frame overlap mirror. Other sources of background are fast neutrons and $\gamma$-particles, which are discriminated by the scintillator detector design and software.

*4.2 Collimation*

As delineated in the previous section, thermal neutrons emerge from an 100 mm × 100 mm area on the surface of the PE moderator. The collimator design takes advantage of this entire area collimating in one dimension to maximize the neutron flux on the sample. At the first glance, a collimator with a length of 1 m seems to be the best option. However, splitting the collimator into two sections introduces flexibility to the setup simplifying future upgrades, including diffraction, radiography, reflectometry, but also a two-dimensional collimation by changing the position of the sections, including a perpendicular alignment. **Figure 9** shows the collimation design consisting of multiple geometrically identical converging channels[31,32] focusing the direct beam onto a single line on the detector.

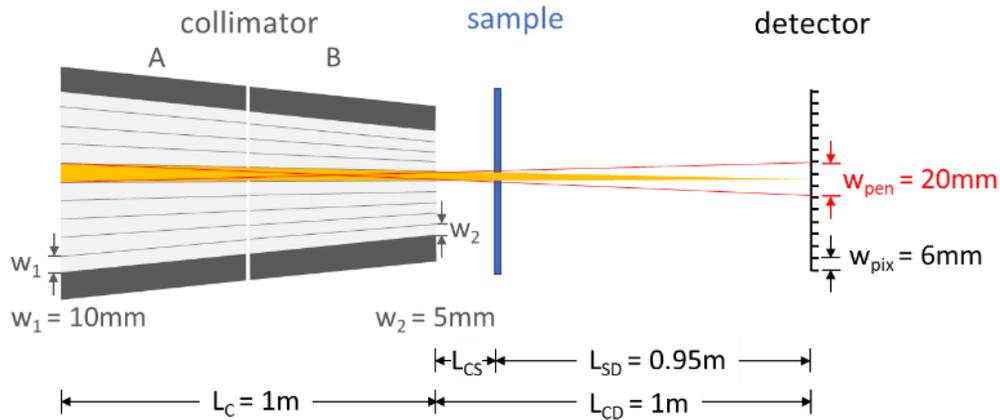

**Figure 9.** Top view of the collimator geometry defining the angular resolution of the SANS instrument, showing the divergence transported through one collimator channel. Ten identical channels, with an entrance width, $w_{1_x} = 10$ mm at the moderator collimated to $w_{2_x} = 5$ mm at the exit are used. The cross-sectional area of the neutrons on the moderator defines the height of the collimator, $h = w_{1_y} = w_{2_y} = 100$ mm, not visible in the top view. The collimator is assembled from 2 pieces of same length (col A and col B), yielding a total length of $L_C = 1$ m. The distance from the end of the collimation to the sample position is 5 cm, and the sample to detector distance is $L_{SD} = 0.95$ cm. The detector pixel area is, $w_{pix_x} \times w_{pix_y} = 6$ mm × 6 mm, and the width of the penumbra originating of one collimation channel is $w_{pen} = 20$ mm.

The width of the penumbra is $w_{pen}$ which is identical for each of the channels can be calculated using geometrical optics,

$$w_{\text{pen}} = \frac{w_2(L_C + L_{CD}) + w_1 L_{CD}}{L_{CD}} \qquad \text{Eq. 9}$$

with the width of the channel at the collimator entrance, $w_1$, the width of the channel at the sample, $w_2$, the length of the collimator, $L_C$, the distance between collimator and detector, $L_{CD}$.[41] The total cross-sectional area of the direct beam at the sample position is defined by the number of channels and the height of the collimator, $h = 100$ mm. For the example, we using $w_1 = 10$ mm, $w_2 = 5$ mm and $L_{CD} = L_C = 1$ m leads to $w_{\text{pen}} = 20$ mm, which defines the theoretically minimum detectable scattering angle,[41] $\theta_{\min} = (0.5 (w_{\text{pen}} + \delta) + w_{\text{pix}})/L_{SD}$. In the experiments, the total beamstop size $B$ should exceed the width of the penumbra $w_{\text{pen}}$ by a small amount $\delta \leq 2$ mm. Therefore, in the vicinity of small angles, $Q_{\min}$ can be calculated by

$$Q_{\min} = \frac{4\pi}{\lambda_{\max}} \theta_{\min} = \frac{4\pi}{\lambda_{\max}} \frac{0.5 (w_{\text{pen}} + \delta) + w_{\text{pix}}}{L_{SD}} \qquad \text{Eq. 10}$$

Using a sample-to-detector distance, $L_{SD} = 0.95$ m and a pixel width, $w_{\text{pix}} = 6$ mm (as detailed in section 4.4) we obtain $\theta_{\min} = 0.018$ rad $\approx 1°$, thus $Q_{\min} \approx 0.011$ Å$^{-1}$ for a maximum wavelength, $\lambda_{\max} = 10$ Å. This $Q_{\min}$ results in an estimated maximum detectable length scale, $\ell_{\max} = \frac{2\pi}{Q_{\min}} \approx 570$ Å. The anticipated size of the detector of 20 cm to each side defines the maximum scattering angle $\theta_{\max} = 0.21$ rad $\approx 12°$, or a $Q_{\max} \approx 2.6$ Å$^{-1}$ for a minimum wavelength, $\lambda_{\min} = 0.5$ Å. Thus, $\ell_{\min} = \frac{2\pi}{Q_{\max}} \approx 2.4$ Å. This resolution range is useful for many sample types.[1-8]

**Frame Overlap Mirror**

While expected to be small, some experiments potentially require suppressing frame-overlap. One traditional method for this is the use of choppers to time-shape the beam. In case of a large slit collimated beam as discussed above, a chopper would need to operate on a rather large beam cross section (about 50 mm x 100 mm, depending on its position along the beam), which is technically challenging and expensive.

An alternative method to address the frame overlap uses V-shaped frame overlap mirrors (FOM)[42] like depicted in **Figure 10** to reflect longer neutron wavelengths out of the direct beam.

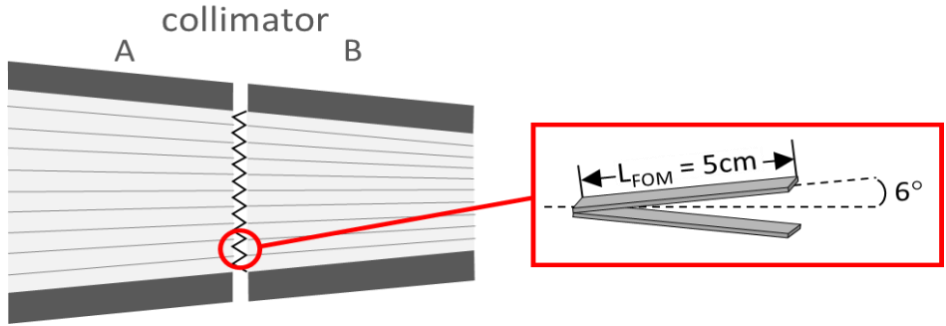

**Figure 10**: Positioning of the frame overlap mirrors (FOM) between collimator A and B. The dimensions and the orientation are given in magnification: The length of one FOM is $L_{FOM} = 50$ mm and the tilting angle is 6°. The width of the mirrors is the same as the height of the collimator $h = 100$ mm.

For neutron supermirrors the critical angle of reflection is given by

$$\gamma^* \approx 0.1 \cdot m \cdot \lambda \qquad \text{Eq. 11}$$

With $m$ being a material parameter increasing with the number of layers of the supermirror.[43] For $m = 4$ and a tilting angle of 6° neutrons wavelengths longer than $\lambda = 15$ Å will be reflected by the mirrors and thus removed from the collimated beam (see **Figure 10**). The tilting angle and $m$-value can be varied to optimize the performance of this component, however for higher $m$-values the manufacturing cost increases rapidly while the transmission decreases.

## 4.3 Sample Position, Flight Tube, and Length of Instrument

Though sample position and flight tube dimensions are limited by the overall design of the instrument, there can be an optional evacuated sample chamber with design choices of connected or separated sample area and flight tube. If the vacuum in these components is separated, the interfaces need to be capped by neutron windows, using materials like Kapton, quartz, and sapphire. Lowering pressure to values between 1 to $10^{-2}$ mbar for a volume of $\pi r^2 l = \pi\, 0.25^2 \text{ m}^2\, 1 \text{ m} \approx 0.2 \text{ m}^3$ can be easily accomplished, e.g., with a rotary vane pump, which requires only little maintenance, like occasional oil changes. However, separation windows introduce background that can in principle be eliminated, simply by avoiding these windows between sample and flight tube and by placing the detector inside the flight tube. The experience at the SAND instrument at IPNS[44] has shown that neutron scattering from a window directly in front of the detector does not pose a problem, since even if thermal neutrons are scattered from it, the resulting Q-value does not change significantly if the detector is placed close behind it. Such a setup simplifies the design and cost of the detector significantly as vacuum feedthroughs and overheating problems of the detector electronics can be avoided.

The total length of the instrument follows the idea to fit it in a typical laboratory. We assume that an overall instrument length of less than 5 m represents a feasible design. Sample to detector distances under $L_{SD} \sim 1$ m are not common for SANS instruments since the spatial resolution at the detector of a few mm impacts the minimum angle accessible and for ToF-SANS the moderator thickness would have a dominant effect on the neutron wavelength resolution, as shown in **Figure 5**. This allows roughly a length of 2 m between moderator and detector with additional space needed between moderator and target, shielding, and maintenance area to access the instrument. A compact design allows high repetition rates and saves cost and floor space, and the angular scattering range can be covered by smaller detector modules.

## 4.4 Area Detector

To accomplish our objective of a reliable, low maintenance instrument, the detector needs to meet several criteria, including: (**1**) A fast response time to handle precise timestamping of neutrons from pulses of the generator at repetition rates of up to 1 kHz. (**2**) Insensitivity or discrimination of fast neutrons and $\gamma$-rays. (**3**) Capsuled to avoid influence of photons from ambient light. (**4**) Robust design, e.g., to avoid damage due to exposure of daylight during operation.

A viable answer to the required specifications seems to be an Anger camera with a Lithium based scintillator coupled to silicon photo multipliers (SiPM)s by a glass plate which acts as a light guide and disperser which allows to pinpoint the origin of the scattering with a fitting algorithm using the photons collected over several pixels simultaneously.[45] These SiPMs have a spatial resolution in the mm range and a high-count rate capability that meets the demands by a high repetition rate of 1 kHz. Additionally, silicon-based photomultipliers are not destroyed by daylight, making operation, and handling easier.[46]

Lithium glass scintillators like GS20[47] are widely used as neutron detectors, because they allow for a very good gamma suppression by pulse height discrimination.[48] The example scintillator GS20 also provides a high, wavelength independent photon yield, which is a concern for non-photon-transparent scintillators when used for a broad neutron wavelength band.

For 2 m distance between moderator and scintillator, the time-of-flight for 0.5 Å neutrons is about 0.25 ms. For optimal wavelength resolution the ideal detector needs a time stamping capability of at least 0.05 ms which is the time resolution limit given by the generator pulse width. Silicon photo multipliers (SiPM) are an affordable technology[46] providing the necessary time and spatial resolution and count rate capability.

A plausible choice for the compact SANS instrument would be to use 4 SiPM panels, each $0.1 \text{ m} \times 0.075 \text{ m}$ arranged in pairs left and right of direct beam, like produced by Nuclear Instruments.[46] A 20 mm gap for the direct beam is sufficient to collect SANS data, without comprising the lowest achievable $Q$.

*4.5 Summary of the Instrument Specifications*

Hereafter, this summary of instrument specifications will concentrate on neutron optics. Additional components like vacuum tubes, a motion stage, and environmental controls for the samples, shielding and a beam dump do not affect the instrument performance and resolution significantly.

- The neutron target and moderator are used as one compact unit and for the simulations in this work they are represented by a flat $100 \text{ mm} \times 100 \text{ mm}$ homogenous surface of pulsed thermalized neutrons variable in frequency up to 200 Hz, the source pulses are 0.05 ms long.
- Following the neutron trajectory downstream from there, a 1m long collimator is placed, with an

initial cross-section of 100 mm × 100 mm. Ten channels, 10 mm wide, converge to 5 mm over the length creating a focus point 1 m downstream the collimator exit.

- The sample cross-section matches the collimator cross section at the exit of 50 mm × 100 mm. Smaller samples can be used reducing the scattered intensity but not the resolution.
- Two kinds of detectors are used, a transmission detector placed in the direct beam at the beam dump and area detector panels which are placed to the left and right of the direct beam at the focus position of the collimator. A pixel size of 6 mm × 6 mm corresponds well with spatial resolution of the SiPM Anger cameras described earlier.

Table 1 gives an overview of the design parameters. The details are described in the sub-chapters for the individual components and Chapter 5.3 shows the simulated performance of this setup.

**Table 1:** Compilation of the design specifications.

1) the SiPM detector covers about 200 mm from the beam center in *x*-direction resulting in $\theta_{\max} = 0.21$ rad with $L_{\text{SD}} = 0.95$ m
2) vertical collimation focusing 10 mm to 5 mm over 1 m resulting in 0.015 rad FWHM, horizontal collimation 100 mm to 100 mm over $L_{\text{C}} = 1$ m results in 0.2 rad FWHM
3) The size of the penumbra is $w_{\text{d}} = 20$ mm; the pixel size is $w_{\text{pix}} = 6$ mm. $\theta_{\min}, Q_{\min}$ are calculated by Eq. 10.

| Design element | Value | |
|---|---|---|
| Source frequency | flexible, up to 1 kHz | |
| Moderator | Polyethylene, room temperature | |
| Thermal flux at sample | $\approx 750$ n/(s cm$^2$) at 200 Hz | |
| Source to sample distance | 1 m | |
| Sample to detector distance | 0.95 m | |
| Collimator to sample distance | 0.05 m | |
| Beamsize at moderator | 0.1 m × 0.1 m | |
| Maximum sample size | 0.1 m × 0.05 m | |
| PSD detectors | 4 panels, each 75 mm × 100 mm | |
| Panel | 1 mm thick GS20 Li-glass scintillator with SiPM detectors | |
| Resolution | 6 mm × 6 mm pixel size, improved be software analysis | |
| Encoding | 12 × 16 spatial pixels with 0.5 ns time binning | |
| Max scattering angle $\theta_{\max}$ | 0.21 rad (f1) | 1) |
| Wavelength band | $0.5 - 10$ Å | |
| Focusing collimator | | |
|       horizontal | 10 mm to 5 mm over 1 m, 0.015 rad FWHM | 2) |
|       vertical | no focusing, 0.2 rad FWHM | 2) |
| $Q_{\min}$ | 0.011 Å$^{-1}$ can be decreased by reducing the source frequency | 3) |
| $Q_{\max}$ | 2.6 Å$^{-1}$ can be increased by installing high angle detector banks | |

## 5 Monte-Carlo Simulations and Data Analysis

In this chapter we use McStas simulations to show a virtual experiment with the previously described setup. McStas[49] is a ray tracing simulation package commonly used to optimize neutron beam optics starting from the moderator. Except for some special materials the interactions of the simulated neutrons with the optical components neglect any energy dependence and the neutrons can be described as rays being reflected, absorbed, or scattered. Any of these interactions are encoded in the direction and amplitude of the rays, which means that each McStas ray does not represent one neutron but a pathway for neutrons with a certain probability $P$. The output from McStas simulations is the summed probability for the events arriving at each detector pixel.

The McStas simulations described in this chapter are intended as example runs to illustrate the potential of a compact SANS. We aim to (**1**) show the figure-of-merit of our instrument by presenting a test case of a standard SANS sample composed of a diluted suspension of spheres with a diameter of 10 nm, (**2**) discuss how to extract and reduce the scattering information and transfer the simulated results into data and (**3**) model the result with SASview in order to show that the scattered data accurately reflects the sample we initially used in the simulation.

### 3.1. Scattering Simulations using McStas

We generate our results using a McStas simulation of the compact instrument consisting of the four key components shown in **Figure 7**, and discussed in chapter 4 in detail:

- a **moderator** 100 mm wide and 100 mm high representing the neutron source in the simulations. To optimize simulation time the output of this components was focused onto a rectangular area 110 cm wide and 200 mm high 2 m downstream, a more divergent output shows no significant effect on the neutron intensity at the area we use in the further analysis.
- a 1m long **collimator** 100 mm high and converging from 100 mm to 50 mm with ten collimation channels for a 1-dimensional horizontal collimation.

- the **sample** is a 100 mm by 50 mm flat rectangle with a 1 mm beam path consisting of diluted and non-interacting 10 nm spheres.
- A flat SANS **detector** 1 m downstream the sample simulated as 300 mm x 300 mm rectangle with 50 channels in in $x$ and $y$ resulting in a 6 mm × 6 mm pixel resolution similar to the SIPM modules.

To keep the simulation time short and simplify the data reduction, the simulation was not run in a time-of-flight mode using discrete wavelength bands (1 Å, 2 Å, 4 Å and 8 Å assuming ± 10% wavelength spread). The neutron scattering for these wavelength intervals will look the same if the data was taken in a TOF simulation but require an analysis and sorting of the events by time in the data reduction, a step which will of course be necessary for real-life experiments but was omitted here for simplicity.

### 3.1.1 Reduction of the simulated SANS data

The following steps were taken to prepare the output from the simulations:

The center of the beam in one dimension $x_{\text{center}}$ was estimated by summing over the 50 channel positions of the detector horizontal coordinate $i$, so $I(i)$ represents the summed intensity for a whole column at each position, the statistical uncertainty $I_{\text{err}}(i)$ for each pixel was summed in the same way and then fitting a Gaussian to this data. Then $I(Q)$ was calculated by transforming the detector $x$-coordinate $i$ into $Q$ approximating the small scattering angles to $|(i - x_{\text{center}}) \cdot 0.006|$, using the distance from the beam center in $x$-direction and the factor 0.006 representing the fact that each detector pixel and thus $i$-value is 6 mm wide in 1 m distance from the sample:

$$I(Q) = \frac{2\pi}{\lambda} \cdot |(i - x_{\text{center}})| \cdot 0.006 \qquad \text{Eq. 12}$$

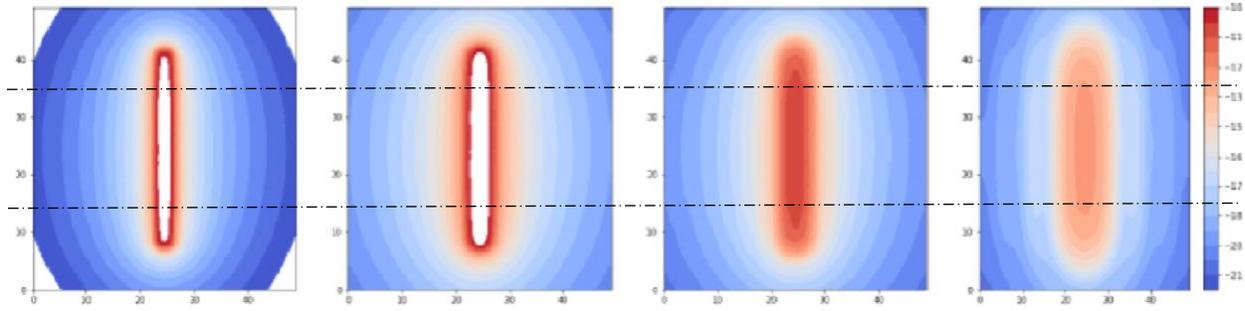

**Figure 11:** Raw data output from the McStas simulation showing the neutrons scattered from the sample from left to right for wavelength bands $1\,\text{Å}, 2\,\text{Å}, 4\,\text{Å}$ and $8\,\text{Å}$ each $\pm\,10\%$. The intensity/probability is displayed on a logarithmic scale. The regions of data enclosed by the dashed lines indicate data used in the reductions, see text for details.

At each $x$-position the McStas output over a 18 pixels high column (along $y$ between the dashed lines in **Figure 11**) was summed. For each wavelength, a scaling factor is applied manually as this simplified data treatment was not scaled on an absolute intensity and the data is written into an output file.

### 3.1.2 Display and Modeling of simulated SANS data

The output from the data reduction of the simulated SANS data can be modeled using SASview.[50] The polydispersity of the spheres in the model was set to 10% as a convenient means to implement the wavelength and thus $Q$-uncertainty into the model. Additionally, the model was smeared by an $0.1\,\text{Å}^{-1}$ high slit-resolution as an estimate for the angular uncertainty, which is also an approximation, as the $Q$-uncertainty changes with wavelength and scattering angle.

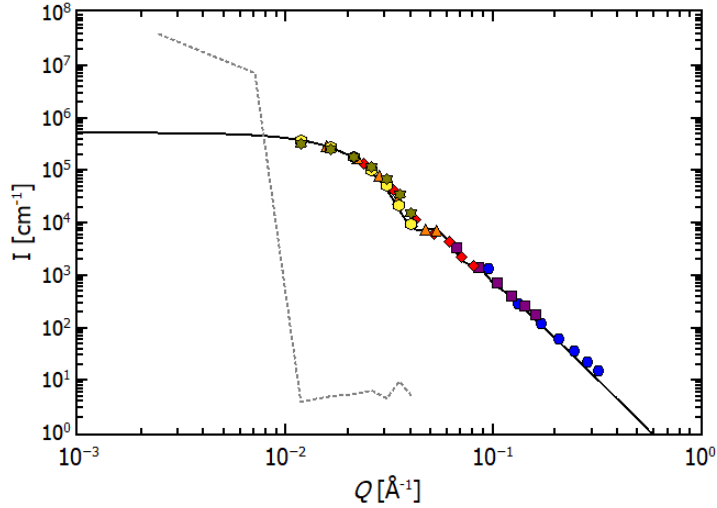

**Figure 12:** The 1D results of the McStas simulations is depicted with blue circles for the 1 Å±10% wavelength band, purple squares for 2 Å±10%, red diamonds for 4 Å±10% and orange triangles up for 8 Å±10%. The grey dotted line represents McStas data without a sample (direct beam), whereas SASview models data for 10 nm diluted spheres smeared by an 0.1 Å$^{-1}$ high slit-resolution with a polydispersity of 10% represented as black line.

At this point the reduction and analysis software is certainly incomplete, data from real experiments needs to be corrected for background, transmission and normalized to an absolute scale and the $Q$-uncertainty must be calculated for each data point. The purple outlier at $Q \approx 0.03$ Å$^{-1}$ is likely related to an offset for the calculated direct beam spot for this wavelength band and other options to fit this need to be explored. Also, we aim to develop a $Q$-resolution treatment in two dimensions (considering $Q_x$ and $Q_y$), so the full detector area becomes usable. However, even using the limited data reduction protocol, **Figure 12** shows that data from our simulation can be modeled by 10 nm spheres and the technique works as expected.

## 3.2 Simulated flux estimates

Here the Adelphi DD109M neutron generator is used as the reference neutron source. Operating at a repetition frequency of 200 Hz will produce an isotropic fast neutron flux of the order of $10^{10}$ n/s.

Moderation by a 50 mm thick PE block at room temperature yields a thermal neutron flux of $\approx 10^7$ n/cm²/s. A rough estimate can be made for an 1 cm² spot on a detector 2 m downstream from the moderator. It covers $\approx 4 \cdot 10^{-5}$ of the solid angle yielding $\approx 10^7$ n/cm²/s $\cdot 4 \cdot 10^{-5} = 400$ n/cm²/s directed flux at the detector for such a setup. Our best estimate for intensity of the newer and stronger versions of Adelphi DD generators, also considering the collimator transmission to be wavelength independent at 70% is a directed flux at the sample of $\approx 750$ n/cm²/s.

Integration of Eq. 3 allows to calculate the relative number of thermal neutrons for the 298K PE moderator, as listed in Table 2.

**Table 2**: Portion of thermal neutrons for selected wavelength bands.

| Wavelength band [Å] | thermal neutrons [%] |
|---|---|
| 0.9 − 1.1 | 16.4 |
| 1.8 − 2.2 | 11.7 |
| 3.6 − 4.4 | 1.36 |
| 7.2 − 8.8 | 0.1 |

To estimate the time for a single SANS sample scan we use the expected flux for the longest wavelength band 750 n/cm²/s $\cdot$ 0.1% = 0.75 n/cm²/s, assuming a 50 cm² cross section sample which scatters 10% of the incoming flux with 90% transmission:

$$I_{\text{est.7.2-8.8Å}} = 0.75 \, \frac{\text{n}}{\text{s} \cdot \text{cm}^2} \cdot 50 \text{ cm}^2 \cdot 0.1 \cdot 0.9 \sim 4 \text{ n}/s \qquad \text{Eq. 13}$$

Under these conditions it takes 7 hours to record 100000 neutrons, which is a rule of thumb for the needed number of neutrons (generating 25 datapoints by averaging data from individual detector pixels, each point will have a statistical uncertainty of $\approx 2\%$).

This estimate represents the time needed to get useful data for the lowest $Q$-value. Using shorter wavelength bands only will allow to perform SANS scans over a reduced $Q$-range within minutes. Cooling of the moderator would increase the flux for the longer neutron wavelengths significantly.

## 6 Conclusions

In this paper we demonstrate the feasibility of a compact SANS setup. While the compact neutron source provides less initial flux than sources at large scale facilities, the flexibility of being the only instrument at the source allows a specialized setup to maximize flux at the sample with a good signal to noise ratio. In this regards the compact SANS design presented in this article has several features to maximize intensity:

- exclusive use of the full moderator surface
- collimation in 1D, slit-smearing the data
- a short overall length allowing a high pulse repetition frequency,
- a larger sample cross section.

Each of these measures gains about one order magnitude in intensity compared to traditional SANS pinhole setups allowing to perform a SANS measurement within a few hours for an ideal sample. As neutron generators are a recent class of neutron sources, we expect that future developments will further improve the compact source strength, which together with the planned cold moderator development will provide significant improvement to the cold neutron flux.

The decision to build the instrument using a slit-smeared resolution poses a challenge for the interpretation of the scattering data. However, it is an often-used option for SAXS experiments and new software tools provide a good handle to tackle this problem and the intensity gains from relaxing the resolution in one dimension are a key factor to make this setup feasible until stronger compact sources re available.

The design of a compact small-angle neutron scattering instrument presented here, creates unique experimental and educational opportunities. The instrument will be particularly useful if samples show very slow structural changes, which may include changes at timescales ranging from days to several years. Introducing a new compact SANS, the experiments can be centered around the timescale given by the sample, independent from the scheduling and proposal constraints of large-scale research facilities. Occasional measurements at traditional SANS instruments can complement measurements at the compact SANS and add to a broader picture, for example allowing studies of kinetics on the time scale of seconds.[15, 51] Fast morphological changes often happen over a limited $Q$-range, but the slower initial and/or equilibrium structure development tends to involve the full $Q$-range. In this way, compact and traditional SANS instruments are complementary, like computers/supercomputers and laboratory and synchrotron SAXS instruments. The independent compact tools add flexibility enabling novel experiments.

This new experiment station can also accommodate samples that might be technically feasible for traditional instruments but cannot be easily shipped or transported there, e.g., samples which are highly sensitive and need a specific environment, hazardous materials or materials which are simply impractical to be shipped.

The newly introduced compact SANS instrument represents a multi-purpose platform that creates several training opportunities. For example, a neutron generator can be operated as a continuous or pulsed source with variable frequency. This allows training of students and staff and testing of instrument components designed for a specific facility more easily.

In conclusion, the compact SANS instrument can potentially play an important role enabling unique experiments, supporting education and training, serve as preparation utility for instruments at high flux sources, and opens the opportunity for users to more easily test sample environments and instrument designs.

## Disclaimer

Specific commercial equipment, instruments, or materials are identified in this paper to foster understanding. Such identification does not imply recommendation or endorsement by the National

Institute of Standards and Technology, nor does it imply that the materials or equipment identified are necessarily the best available for the purpose.

## References

1. Gerstl, C.; Schneider, G. J.; Pyckhout-Hintzen, W.; Allgaier, J.; Willbold, S.; Hofmann, D.; Disko, U.; Frielinghaus, H.; Richter, D., Chain Conformation of Poly(alkylene oxide)s Studied by Small-Angle Neutron Scattering. *Macromolecules* **2011,** *44* (15), 6077-6084.
2. Gerstl, C.; Brodeck, M.; Schneider, G. J.; Su, Y.; Allgaier, J.; Arbe, A.; Colmenero, J.; Richter, D., Short and Intermediate Range Order in Poly(alkylene oxide)s. A Neutron Diffraction and Molecular Dynamics Simulation Study. *Macromolecules* **2012,** *45* (17), 7293-7303.
3. Schneider, G. J.; Vollnhals, V.; Brandt, K.; Roth, S. V.; Goritz, D., Correlation of mass fractal dimension and cluster size of silica in styrene butadiene rubber composites. *J Chem Phys* **2010,** *133* (9), 094902.
4. Sternhagen, G. L.; Gupta, S.; Zhang, Y.; John, V.; Schneider, G. J.; Zhang, D., Solution Self-Assemblies of Sequence-Defined Ionic Peptoid Block Copolymers. *Journal of the American Chemical Society* **2018,** *140* (11), 4100-4109.
5. De Mel, J. U.; Gupta, S.; Perera, R. M.; Ngo, L. T.; Zolnierczuk, P.; Bleuel, M.; Pingali, S. V.; Schneider, G. J., Influence of external NaCl salt on membrane rigidity of neutral DOPC vesicles. *Langmuir* **2020**.
6. Gupta, S.; De Mel, J. U.; Perera, R. M.; Zolnierczuk, P.; Bleuel, M.; Faraone, A.; Schneider, G. J., Dynamics of Phospholipid Membranes beyond Thermal Undulations. *The Journal of Physical Chemistry Letters* **2018,** *9* (11), 2956-2960.
7. Gupta, S.; Chatterjee, S.; Zolnierczuk, P.; Nesterov, E. E.; Schneider, G. J., Impact of Local Stiffness on Entropy Driven Microscopic Dynamics of Polythiophene. *Scientific reports* **2020,** *10* (1), 1-10.
8. Gilbert, D. A.; Grutter, A. J.; Neves, P. M.; Shu, G.-J.; Zimanyi, G.; Maranville, B. B.; Chou, F.-C.; Krycka, K.; Butch, N. P.; Huang, S.; Borchers, J. A., Precipitating ordered skyrmion lattices from helical spaghetti and granular powders. *Physical Review Materials* **2019,** *3* (1), 014408.
9. Mühlbauer, S.; Honecker, D.; Périgo, É. A.; Bergner, F.; Disch, S.; Heinemann, A.; Erokhin, S.; Berkov, D.; Leighton, C.; Eskildsen, M. R.; Michels, A., Magnetic small-angle neutron scattering. *Reviews of Modern Physics* **2019,** *91* (1), 015004.
10. Mühlbauer, S.; Binz, B.; Jonietz, F.; Pfleiderer, C.; Rosch, A.; Neubauer, A.; Georgii, R.; Böni, P., Skyrmion Lattice in a Chiral Magnet. *Science* **2009,** *323* (5916), 915-919.
11. UCANS. www.ucans.org.
12. Kiyanagi, Y., JCANS network of compact neutron facilities in Japan. *The European Physical Journal Plus* **2016,** *131* (5), 132.
13. Thomas, B.; Thomas, G.; Sebastian, S.; Christiane, A.-S.; Frédéric, O.; Alain, M., Low energy accelerator-driven neutron facilities—A prospect for a brighter future for research with neutrons. *Neutron News* **2020,** *31* (2-4), 13-18.
14. Laxdal, R.; Maharaj, D.; Abbaslou, M.; Tun, Z.; Banks, D.; Gottberg, A.; Marchetto, M.; Rodriguez, E.; Yamani, Z.; Fritzsche, H.; Rogge, R.; Pan, M.; Kester, O.; Marquardt, D., A Prototype Compact Accelerator-based Neutron Source (CANS) for Canada. *Journal of Neutron Research* **2021,** *23*, 99-117.
15. Gupta, S.; Bleuel, M.; Schneider, G. J., A new ultrasonic transducer sample cell for in situ small-angle scattering experiments. *Review of Scientific Instruments* **2018,** *89* (1), 015111.
16. Radulescu, A.; Szekely, N. K.; Appavou, M.-S.; Pipich, V.; Kohnke, T.; Ossovyi, V.; Staringer, S.; Schneider, G. J.; Amann, M.; Zhang-Haagen, B.; Brandl, G.; Drochner, M.; Engels, R.; Hanslik,

R.; Kemmerling, G., Studying Soft-matter and Biological Systems over a Wide Length-scale from Nanometer and Micrometer Sizes at the Small-angle Neutron Diffractometer KWS-2. **2016,** (118), e54639.
17. Wu, J.; Xu, S.; Han, C. C.; Yuan, G., Controlled drug release: On the evolution of physically entrapped drug inside the electrospun poly(lactic-co-glycolic acid) matrix. *Journal of Controlled Release* **2021,** *331*, 472-479.
18. Weber, J.; Cheshire, M. C.; Bleuel, M.; Mildner, D.; Chang, Y.-J.; Ievlev, A.; Littrell, K. C.; Ilavsky, J.; Stack, A. G.; Anovitz, L. M., Influence of microstructure on replacement and porosity generation during experimental dolomitization of limestones. *Geochimica et Cosmochimica Acta* **2021,** *303*, 137-158.
19. Pelletier, S.; Jabali, O.; Laporte, G.; Veneroni, M., Battery degradation and behaviour for electric vehicles: Review and numerical analyses of several models. *Transportation Research Part B: Methodological* **2017,** *103*, 158-187.
20. Amorim Júnior, N. S.; Silva, G. A. O.; Dias, C. M. R.; Ribeiro, D. V., Concrete containing recycled aggregates: Estimated lifetime using chloride migration test. *Construction and Building Materials* **2019,** *222*, 108-118.
21. Jiří, V., Polymer degradation: a short review. *Chemistry Teacher International* **2021,** *3* (2), 213--220.
22. Sazali, N.; Ibrahim, H.; Jamaludin, A. S.; Mohamed, M. A.; Salleh, W. N. W.; Abidin, M. N. Z., Degradation and stability of polymer: A mini review. *IOP Conference Series: Materials Science and Engineering* **2020,** *788* (1), 012048.
23. Zatoń, M.; Rozière, J.; Jones, D. J., Current understanding of chemical degradation mechanisms of perfluorosulfonic acid membranes and their mitigation strategies: a review. *Sustainable Energy & Fuels* **2017,** *1* (3), 409-438.
24. Carpenter, M.; Faber, J., Jnr, Design study of a time-of-flight small-angle diffractometer for a pulsed neutron source. *Journal of Applied Crystallography* **1978,** *11* (5), 464-465.
25. Sears, V. F.; Campbell, R. J.; Sears, R. S. A. E. C. L. V. F., *Neutron Optics: An Introduction to the Theory of Neutron Optical Phenomena and Their Applications*. Oxford University Press: 1989.
26. Atkins, P. W., *Physikalische Chemie*. 2 ed.; Weinheim: New York, Basel, Cambridge, Tokyo, VCH, 1996.
27. Kisi, E. H.; Howard, C. J., *Applications of Neutron Powder Diffraction*. Oxford University Press: New York, 2008.
28. Mildner, D. F. R.; Carpenter, M. J., Optimization of the experimental resolution for small-angle scattering. *Journal of Applied Crystallography* **1984,** *17* (4), 249--256.
29. Seeger, P. A., Optimization of geometric resolution in small-angle scattering. *Nuclear Instruments and Methods* **1980,** *178* (1), 157-161.
30. Mildner, D. F. R.; Carpenter, J. M., Resolution of small-angle scattering with Soller collimation. *Journal of Applied Crystallography* **1987,** *20* (5), 419-424.
31. Hjelm, R., Jnr, The resolution of TOF low-Q diffractometers: instrumental, data acquisition and reduction factors. *Journal of Applied Crystallography* **1988,** *21* (6), 618-628.
32. Lake, J. A., An iterative method of slit-correcting small angle X-ray data. *Acta Crystallographica* **1967,** *23* (2), 191-194.
33. Feigin, L. A.; Svergun, D. I., *Structure Analysis by Small-Angle X-ray and Neutron Scattering*. Plenum: New York, 1987.
34. Pauw, B. R., Everything SAXS: small-angle scattering pattern collection and correction. *Journal of Physics: Condensed Matter* **2013,** *25* (38), 383201.
35. Radia, I.; Rep, Y., Neutron Generators for Analytical Purposes. *IAEA radiation technology reports series, ISSN* **2012**, 2225-8833.
36. Sahiner, H.; Norris, E. T.; Bugis, A. A.; Liu, X., Improved shielding design with an accelerated Monte Carlo simulation for a neutron generator at Missouri S&T. *Progress in Nuclear Energy* **2017,** *97*, 123-132.

37. Ericsson, G., Advanced Neutron Spectroscopy in Fusion Research. *Journal of Fusion Energy* **2019,** *38* (3), 330-355.
38. Vainionpaa, J. H.; Chen, A. X.; Piestrup, M. A.; Gary, C. K.; Jones, G.; Pantell, R. H., Development of high flux thermal neutron generator for neutron activation analysis. *Nuclear Instruments and Methods in Physics Research Section B: Beam Interactions with Materials and Atoms* **2015,** *350*, 88-93.
39. Vainionpaa, J. H.; Harris, J. L.; Piestrup, M. A.; Gary, C. K.; Williams, D. L.; Apodaca, M. D.; Cremer, J. T.; Ji, Q.; Ludewigt, B. A.; Jones, G., High yield neutron generators using the DD reaction. *AIP Conference Proceedings* **2013,** *1525* (1), 118-122.
40. Cheng, C.; Wei, Z.; Hei, D.; Jia, W.; Sun, A.; Li, J.; Cai, P.; Zhao, D.; Shan, Q.; Ling, Y., Design of a PGNAA facility using D-T neutron generator for bulk samples analysis. *Nuclear Instruments and Methods in Physics Research Section B: Beam Interactions with Materials and Atoms* **2019,** *452*, 30-35.
41. Crawford, R.; Thiyagarajan, P.; Epperson, J.; Trouw, F.; Kleb, R.; Wozniak, D.; Leach, D. *The new small-angle diffractometer SAND at IPNS*; Argonne National Lab., IL (United States): 1995.
42. Böni, P.; Schanzer, C.; Schneider, M., Wide-angle transmission analyzer for polarized neutrons using equiangular spirals. *Nuclear Instruments and Methods in Physics Research Section A: Accelerators, Spectrometers, Detectors and Associated Equipment* **2020,** *966*, 163858.
43. Turchin, V. F., Diffraction of slow neutrons by stratified systems. *Soviet Atomic Energy* **1967,** *22* (2), 124-125.
44. Thiyagarajan, P.; Crawford, R. K.; Epperson, J. E.; Trouw, F.; Kleb, R.; Wozniak, D.; Leach, D., The performance of the small-angle diffractometer, SAND at IPNS. *Acta Crystallographica* **1998,** *52*, C36 - C36.
45. Morozov, A.; Marcos, J.; Margato, L.; Roulier, D.; Solovov, V., SiPM-based neutron Anger camera with auto-calibration capabilities. *Journal of Instrumentation* **2019,** *14* (03), P03016-P03016.
46. www.nuclearinstruments.eu/dt5550afes.html.
47. www.scintacor.com/products/6-lithium-glass.
48. Spowart, A. R., Measurement of the gamma sensitivity of granular and glass neutron scintillators and films. *Nuclear Instruments and Methods* **1970,** *82*, 1-6.
49. Lefmann, K.; Nielsen, K., McStas, a general software package for neutron ray-tracing simulations. *Neutron News* **1999,** *10* (3), 20-23.
50. www.sasview.org/docs/user/qtgui/Perspectives/Fitting/resolution.html.
51. Lund, R.; Willner, L.; Richter, D., Kinetics of Block Copolymer Micelles Studied by Small-Angle Scattering Methods. In *Controlled Polymerization and Polymeric Structures: Flow Microreactor Polymerization, Micelles Kinetics, Polypeptide Ordering, Light Emitting Nanostructures*, Abe, A.; Lee, K.-S.; Leibler, L.; Kobayashi, S., Eds. Springer International Publishing: Cham, 2013; pp 51-158.